\documentclass[12pt]{article}

\usepackage{float}
\usepackage{graphicx}
\usepackage{amsmath,amsfonts,amssymb}

\newtheorem{thm}{Theorem}[section]

\date{ }
\begin{document}

\begin{center}

{\bf \large Reliability of MST identification in correlation-based market networks}

\vskip 2mm
Kalyagin V.A.\footnote{Corresponding author. E-mail: vkalyagin@hse.ru}, Koldanov A.P., Koldanov P.A.

\vskip 2mm
National Research University Higher School of Economics, \\ 
Laboratory of Algorithms and Technologies for Network Analysis, \\
136 Rodionova street, 603093, Nizhny Novgorod, Russia

%\address[latna1]{National Research University Higher School of Economics, \\ Laboratory of Algorithms and
%Technologies for Network Analysis, 136 Rodionova street, 603093, Nizhny Novgorod, Russia}

\end{center}

\begin{abstract}
Maximum spanning tree (MST) is a popular tool in market network analysis. Large number of publications are devoted to the MST calculation and it's interpretation for particular stock markets. However, much less attention is payed in the literature to the analysis of uncertainty of obtained results. In the present paper we suggest a general framework to measure uncertainty of MST identification. We study uncertainty in the framework of the concept of random variable network (RVN).  We consider different correlation based networks in the large class of elliptical distributions.  We show that true MST is the same in three networks: Pearson correlation network, Fechner correlation network, and Kendall correlation network. We argue that among different measures of uncertainty the FDR (False Discovery Rate) is the most appropriated for MST identification. We investigate FDR of Kruskal algorithm for MST identification  and show that reliability of MST identification is different in these three networks. In particular,  for Pearson correlation network the FDR essentially depends on distribution of stock returns. We prove that for market network with Fechner correlation the FDR is non sensitive to the assumption on stock's return distribution. Some interesting phenomena are discovered for Kendall correlation network. Our experiments show that FDR of Kruskal algorithm for MST identification in Kendall correlation network weakly depend on distribution and at the same time the value of FDR is almost the best in comparison with MST identification in other networks. These facts are important in practical applications.  

\end{abstract}
  

\section{Introduction}
Network models of stock market have attracted a large attention in theoretical and applied
research. Different graph structures related with stock market network are considered in the literature \cite{Boginski_2006}. One of such graph structure, maximum spanning tree (MST), is a popular tool in market network analysis. Many papers are devoted to the use of MST for particular stock markets (see recent papers  \cite{Walle_2018}, \cite{Sensoya_2014}, \cite{Wang_2012} and exhaustive bibliography in \cite{Marti_Arxive}). However, much less attention is payed in the literature to the estimation of uncertainty of obtained results. One particular way to measure uncertainty is related with bootstrap technique applied to the observed data \cite{Tumminello_2007}. In the present paper we suggest a general theoretical framework to measure uncertainty of MST identification. We study uncertainty in the framework of the concept of random variable network \cite{Kalyagin_2020}. 

Random variable network (RVN) is a pair $(X,\gamma)$, where $X=(X_1,X_2,\ldots,X_N)$ is a random vector and $\gamma$ is a measure of similarity between pairs of random variables. This concept allows to introduce the {\it true MST} associated with RVN. We call true MST the maximum spanning tree in the complete weighted graph $(V,\Gamma)$, where $V=\{1,2,\ldots,N\}$ is the set of nodes (vertices), and $\Gamma=(\gamma_{i,j})$ is the matrix of weights, $\gamma_{i,j}=\gamma(X_i,X_j)$, $i,j=1,2,\ldots,N$, $i \neq j$, $\gamma_{i,j}=0$ for $i=j$. To model distribution of the vector $X$ we use a large class of elliptical distributions, which is widely used in applied finance \cite{Gupta_2013}. To measure similarity between stock's we consider different correlation networks for the stock's returns: Pearson correlation network, Fechner correlation network, and Kendall correlation network. Pearson correlation is most used in market network analysis. We show in the paper that for elliptical distributions the true  MST in Fechner and Kendall correlation networks are the same as the true MST in Pearson correlation network for Gaussian distribution. This fact gives a theoretical basis for correct comparison of uncertainty of MST identification algorithms in different networks.

Uncertainty of MST identification in our setting is related with the difference between true MST and MST identified from observations. To assess uncertainty of MST identification we analyze different error rates known in multiple testing and binary classification. We argue that the most appropriate error rate for MST identification is the well known False Discovery Rate (FDR). In our case FDR is the proportion of false edges (non correctly identified edges) in MST. We investigate FDR of Kruskal algorithm for MST identification  and show that reliability of MST identification is different in three correlation networks. We emphasize that for Pearson correlation network the FDR essentially depends on distribution of stock returns. We prove that for  Fechner correlation network the FDR is non sensitive to the assumption on stock's return distribution. New and surprising phenomena are discovered for   Kendall correlation network. Our experiments show that FDR of Kruskal algorithm for MST identification in Kendall correlation network weakly depend on distribution and at the same time the value of FDR is almost the best in comparison with MST identification in other networks. These facts are important in practical applications.  
 
The paper is organized as follows. In Section \ref{Basic definitions and notations} we present necessary definitions and notations. In Section \ref{Connection} we prove that MST is the same in three correlation networks for a large class of elliptical distributions. In Section \ref{Uncertainty of MST} we discuss measures of uncertainty of MST identification. The Section \ref{Kruskal algorithm for MST identification} is devoted to the description of Kruskal algorithms for MST identification in different correlation networks. In Section \ref{Robustness of Kruskal algorithm} we prove robustness of Kruskal algorithm of MST identification in Fechner correlation network. In Section \ref{Reliability of Kruskal algorithm} we present the results of numerical investigation of reliability of Kruscal algorithm in different correlation networks. The Section \ref{Concluding remarks} summarizes the main results of the paper and discusses a further research. 

\section{Basic definitions and notations.}\label{Basic definitions and notations}

Random variable network is a pair $(X,\gamma)$, where $X=(X_1,\ldots,X_N)$ is a random vector, and $\gamma$ is a pairwise measure of similarity between random variables. 
One can consider different random variable networks associated with different distributions of the random vector $X$ and different measures of similarity $\gamma$. For example, the Gaussian Pearson correlation network is the random variable network, where $X$ has a multivariate Gaussian distribution and $\gamma$ is the Pearson correlation. On the same way one can consider the  Gaussian partial correlation network, the Gaussian Kendall correlation network, the Student Pearson correlation network and so on. 

The random variable network generates a network model. Network model for random variable network $(X, \gamma)$ is the complete weighted graph $(V,\Gamma)$ with $N$ nodes , where $V=\{1,2,\ldots,N\}$ is the set of nodes, $\Gamma=(\gamma_{i,j})$ is the matrix of weights, $\gamma_{i,j}=\gamma(X_i,X_j)$. The spanning tree in the network model $(V,\Gamma)$ is a connected graph $(V,E)$ without cycles. Weight of the  spanning tree $(V,E)$ is the sum of weights of its edges $\sum_{(i,j) \in E} \gamma_{i,j}$. Maximum spanning tree (MST) is the spanning tree with maximal weight. In what follows we consider MST as unweighted  graph. MST obtained in this way will be called {\it true MST} or {\it MST in true network model}. There are known many algorithms to calculate the minimum spanning tree in an undirected weighted graph \cite{Gross_2006}. All of them can be easily transformed onto algorithms to calculate the maximum spanning tree. In this paper we use classical Kruscal algorithm:

\vskip 2mm
\noindent 
{\bf Kruskal algorithm for calculation of the true MST:} the Kruskal algorithm calculates the collection of edges $MST$ of maximum spanning tree in the network model $(V,\Gamma)$ by the following steps
\begin{itemize}
\item Sort the edges of the complete weighted graph $(V,\Gamma)$ into decreasing order by weights $\gamma_{i, j}$. 
\item Add the first edge to $MST$.
\item Add the next edge to $MST$ if and only if it does not form a cycle in the current $MST$. 
\item If $MST$ has $(N-1)$ edges, stop and output $MST$ . Otherwise go to the previous step.
\end{itemize}

We consider three correlation networks, Pearson correlation network, Fechner correlation network, and Kendall correlation network with elliptical distribution of the vector $X$.
Pearson correlation network is a random variable network with Pearson correlation as the measure of similarity $\gamma=\gamma^P$ 
\begin{equation}\label{Pearson measure}
\gamma^P_{i,j}=\gamma^P(X_i,X_j)=\frac{Cov(X_i,X_j)}{\sqrt{Cov(X_i,X_i)}\sqrt{Cov(X_j,X_j)}}
\end{equation}
Fechner correlation network is a random variable network with Fechner correlation as the measure of similarity $\gamma=\gamma^{Fh}=2 \gamma^{Sg}-1$
where $\gamma^{Sg}$ is so-called sign similarity
\begin{equation}\label{Sign similarity}
\gamma^{Sg}_{i,j}=\gamma^{Sg}(X_i,X_j)=P\{(X_i-E(X_i))(X_j-E(X_j))>0\} 
\end{equation}
\begin{equation}\label{Fechner correlation}
\gamma^{Fh}_{i,j}=2\gamma^{Sg}_{i,j}-1
\end{equation}
Kendall correlation network is a random variable network with Kendall correlation as the measure of similarity $\gamma=\gamma^{Kd}$
\begin{equation}\label{Kendall correlation}
\gamma^{Kd}_{i,j}=\gamma^{Kd}(X_i,X_j)=2P\{(X_i^{(1)}-X_i^{(2)})(X_j^{(1)}-X_j^{(2)})>0\}-1
\end{equation}
where $(X_i^{(1)}, X_j^{(1)})$, $(X_i^{(2)}, X_j^{(2)})$ are two independent random vectors with the same distribution as the vector $(X_i,X_j)$ (see \cite{Kruskal_1958}). 

Random vector $X$ belong to the class of  elliptically contoured distributions (elliptical distributions) if its  density function has the form  \cite{Anderson_2003}:
\begin{equation}\label{density_of_elliptical_distribution} 
f(x; \mu, \Lambda)=|\Lambda|^{-\frac{1}{2}}g\{(x-\mu)'\Lambda^{-1}(x-\mu)\}
\end{equation}
where $\Lambda=(\lambda_{i,j})_{i,j=1,2,\ldots,N}$ is positive definite symmetric matrix, $g(x)\geq 0$, and 
$$
\int_{-\infty}^{\infty}\ldots\int_{-\infty}^{\infty}g(y'y)dy_1 dy_2\cdots dy_N=1
$$
This class includes in particular multivariate Gaussian distribution  
$$
f_{Gauss}(x)=\frac{1}{(2\pi)^{N/2}|\Lambda|^{\frac{1}{2}}}e^{-\frac{1}{2}(x-\mu)'\Lambda^{-1}(x-\mu)}
$$
and multivariate Student distribution with $\nu$ degree of freedom
$$
f_{Student}(x)=\frac{\Gamma\left(\frac{\nu+N}{2}\right)}{\Gamma\left(\frac{\nu}{2}\right)\nu^{N/2}\pi^{N/2}}|\Lambda|^{-\frac{1}{2}}\left[1+\frac{(x-\mu)'\Lambda^{-1}(x-\mu)}{\nu}\right]^{-\frac{\nu+N}{2}}
$$
The class of elliptical distributions is a natural generalization of the class of Gaussian distributions. Many properties of Gaussian distributions have analogs for elliptical distributions, but this class is much larger, in particular it includes distributions with heavy tails. For detailed investigation of elliptical distributions see  \cite{Fang_1990}, \cite{Anderson_2003}, \cite{Gupta_2013}. It is known that if $E(X)$ exists then $E(X)=\mu$. One important property of elliptical distributions is the connection between covariance matrix of the vector $X$ and the matrix $\Lambda$. Namely, if covariance matrix exists one has
\begin{equation}\label{Covariance_Lambda}
\sigma_{i,j}=Cov(X_i, X_j)=C \cdot \lambda_{i,j}
\end{equation}
where 
$$
C=\frac{2\pi^{\frac{1}{2}N}}{\Gamma(\frac{1}{2}N)}\int_0^{+\infty}r^{N+1}g(r^2)dr
$$
In particular, for Gaussian distribution one has $Cov(X_i, X_j)=\lambda_{i,j}$. For multivariate Student distribution with $\nu$ degree of freedom ($\nu>2$) one has $\sigma_{i,j}=\nu/(\nu-1)\lambda_{i,j}$.

\section{Connection between random variable networks}\label{Connection}
There is a connection between  three networks for the vector $X$ with elliptical distribution with the same matrix $\Lambda$. Let $\Lambda$ be a fixed  positive definite matrix of dimension $(N \times N)$. Denote by $K(\Lambda)$ the class of distributions such that its  density function has the form (\ref{density_of_elliptical_distribution}). The following statement holds.
\begin{thm}
Let $X \in K(\Lambda)$. If covariance matrix of $X$ exists, then the true MST in Pearson, Fechner, and Kendall correlation networks is the same for any network and any distribution of the vector $X$. This MST coincides with the true MST for multivariate Gaussian distribution with the covariance matrix $\Lambda$.   
\end{thm}

\noindent
{\bf Proof.}  Let $X$ be a random vector with  elliptical distribution (\ref{density_of_elliptical_distribution}) with the matrix $\Lambda$ and the function $g(u)$. The relation (\ref{Covariance_Lambda}) implies that 
$$
\gamma^P(X_i,X_j)=\displaystyle \frac{\lambda_{i,j}}{\sqrt{\lambda_{i,i}\lambda_{j,j}}},
$$
that is $\gamma^P(X_i,X_j)$ does not depend on the function $g(u)$ and are defined by the matrix $\Lambda$ only. We will prove that this is true for Fechner and Kendall correlations too. This fact is proved for the sign similarities $\gamma_{i,j}^{Sg}$ in \cite{Kalyagin_2017}, Lemma 1 and Lemma 2. Therefore it is true for Fechner correlations $\gamma^{Fh}_{i,j}=2\gamma_{i,j}^{Sg}-1$. Moreover it is proved in \cite{Kalyagin_2017} that
$$
\gamma_{i,j}^{Sg}=\displaystyle \frac{1}{2}+\frac{1}{\pi}\arcsin (\gamma^P_{i,j})
$$
For Kendall correlations consider two  independent random vectors $X^{(1)}$, $X^{(2)}$  with the same distribution as the vector $X$. It can be easy proved that in this case the random vector $(X^{(1)}-X^{(2)})$ has elliptical distribution \cite{Lindskog_2003}. Calculation of the covariance matrix for this vector implies  
$$
Cov(X^{(1)}_i-X^{(2)}_i, X^{(1)}_j-X^{(2)}_j)=2 Cov(X_i,X_j)=2C\lambda_{i,j}
$$
Therefore
$$
\gamma_{i,j}^{Kd}=2\gamma^{Sg}(X^{(1)}_i-X^{(2)}_i,X^{(1)}_j-X^{(2)}_j)-1=\displaystyle \frac{2}{\pi}\arcsin(\frac{\lambda_{i,j}}{\sqrt{\lambda_{i,i}\lambda_{j,j}}})
$$
It implies that Kendall correlations don't depend on the function $g(u)$. 
Moreover the following relations  hold for any distribution from the class $K(\Lambda)$: 
$$
\gamma^{Fh}_{i,j}=\gamma^{Kd}_{i,j}=\displaystyle \frac{2}{\pi}\arcsin(\gamma^P_{i,j}).
$$ 

To calculate the true MST in each of three networks one can use Kruskal algorithm. The first step of the algorithm is to sort the edges of the complete weighted graph $(V,\Gamma)$ into decreasing order by weights $\gamma_{i, j}$. Note, that $\gamma^{Fh}_{i,j}$, $\gamma^{Kd}_{i,j}$ are obtained from  $\gamma^P_{i,j}$ by increasing function. It means that the first step of the Kruskal algorithm will give the same edge ordering for all three networks. Next steps of the algorithm depends only on this ordering  and does not depend on a particular values of the weights of edges. Therefore the true maximum spanning tree (true MST) is the same in all networks for any distribution of the vector $X \in K(\Lambda)$. True MST for multivariate Gaussian distribution with the covariance matrix $\Lambda$ is a particular case of such MST.   

This statement gives a basis for a correct comparison of reliability of Kruskal algorithm for MST identification in different correlation networks. 

\noindent
{\bf Remark:} The relation between Kendall and Pearson correlation for elliptical distributions is known \cite{Fang_2002}, \cite{Lindskog_2003}. We give here a 
sketch of proof by the sake of completeness.

\section{Uncertainty of MST identification in random variable network}\label{Uncertainty of MST}
The main problem under discussion in this paper is a reliability of the identification of the true MST from observations. Let $(X,\gamma)$ be a random variable network and $(V,\Gamma)$ be the associated network model. True maximum spanning tree (true MST) is the spanning tree in  $(V,\Gamma)$ with maximal weight.  Let $X(t)$, $t=1,2,\ldots, n$ be a sample from distribution of $X$. Denote by $x(t)$ observed value of the random vector $X(t)$. Sample space is defined by the matrices $x=(x_j(t)) \in R^{N \times n}$.  We define the decision space as the space of all adjacency matrices $S$ of the spanning trees in $(V,\Gamma)$:
$$
{\cal{D}}=\{S : \ S \in R^{N \times N}, S \ \mbox{ is adjacency matrix for a spanning tree in} \ (V,\Gamma) \}
$$
Any MST identification algorithm $\delta=\delta(x)$ is a map from the sample space $R^{N \times n}$ to the decision space $\cal{D}$. 
Quality of an identification algorithm $\delta$ is related with the difference between true maximum spanning tree  and the spanning tree given by $\delta$ which can be evaluated by a loss function $w(S,Q)$ where $S=(s_{i,j})$ is the true decision and $Q=(q_{i,j})$ is the decision given by $\delta$. 
Uncertainty of an identification algorithm $\delta$ is then measured by the expected value of the loss function, which is known as the risk function
\begin{equation}\label{risk_function}
Risk(S; \delta)= \sum_{Q \in \cal{D}} w(S,Q)P(\delta=Q)
\end{equation}
The choice of the loss function is an important point for uncertainty evaluation. To discuss an appropriate choice of the loss function for MST identification we consider the following tables familiar in binary classification. Table \ref{s_ij_q_ij} illustrates Type I and Type II errors for the individual edge $(i,j)$. It represents all possible cases for different values of $s_{i,j}$ and $q_{i,j}$. Value $0$ means that the edge $(i,j)$ is not included in the MST, 
value $1$ means that the edge $(i,j)$ is  included in the MST. We associate the case $s_{i,j}=0$, $q_{i,j}=1$ with Type I error (false edge inclusion), and we associate the case $s_{i,j}=1$, $q_{i,j}=0$ with Type II error (false edge non inclusion). 

\begin{table}[h!]
\begin{center}
  \begin{tabular}{|c|c|c|}
\hline
  &  &  \\
$q_{i,j}$   $s_{i,j}$ & 0 & 1 \\
  &  &  \\
\hline
  &  &  \\
0 & \ edge is not included correctly & Type II error \  \\
  &  &  \\
\hline
  &  &  \\
1 & \ Type I error & edge is included correctly \ \\
  &  &  \\
\hline
  \end{tabular}
	\end{center}
 \caption{Type I (false edge inclusion) and Type II (false edge exclusion) errors for the edge $(i,j)$}\label{s_ij_q_ij}
 \end{table}

Table \ref{TN_TP} represents the numbers of Type I errors (False Positive), number of Type II errors (False Negative), and numbers of correct decisions (True Positive and True Negative). This table has a specific properties for the numbers of errors in MST identification. First, number of edges in any spanning tree is equal to $(N-1)$, 
that is $FP+TP=N-1$, $FP+TN=M-(N-1)$, where $M=C^2_N$. Second, one false included edge implies one false excluded edge and vice versa, that is $FP=FN$. In addition one has $0 \leq FP \leq N-1$, $M-2(N-1) \leq TN \leq M-(N-1)$.

 \begin{table}[h!]
\begin{center}
  \begin{tabular}{|c|c|c|c|}
\hline
  &  &  & \\
$Q \ S$ & 0 in $S$ & 1 in $S$ & Total\\
  &  &  & \\
\hline
  &  &  & \\
0 in $Q$ &  TN & FN  &  number of 0 in $Q$ \\
  &  &  & \\
\hline
  &  &  & \\
1 in $Q$ & FP & TP & number of 1 in $Q$ \ \\
  &  &  & \\
\hline
  &  &  & \\
Total & number of 0 in $S$ & number of 1 in $S$ & $N(N-1)/2$ \\
  &  &  & \\
\hline
  \end{tabular}
\end{center}
 \caption{Numbers of Type I and Type II errors for MST identification}\label{TN_TP}
 \end{table}

Now we discuss the choice of the loss and risk functions appropriate for the MST identification by observations. The most simple loss function is 
$$
w_{Simple}(S,Q)= 
\left \{
\begin{array}{lll}
1 & if & S \neq Q \\
0 & if & S = Q \\
\end{array}
\right.
$$
The associated risk is the probability of the false decision $Risk(S; \delta)=P(\delta(x) \neq S)$. For MST identification it is the same as 
FWER (Family Wise Error Rate), known in multiple hypotheses testing. FWER is the probability of at least one Type I error \cite{Hochberg_1987}, \cite{Bretz_2011}, \cite{Lehmann_2005}. The true MST is correctly identified if and only if $FP=FN=0$. This measure of uncertainty takes into account only the fact of correct identification of MST (no errors) and it does not take into account the number of errors. Moreover, one can show by simulations, that the probability of correct decision for MST identification is very small even if the number of observations is big \cite{Kalyagin_2014_1}. 

Another error rates such as Conjunctive Power (CPOWER) and Disjunctive Power (DPOWER), known in multiple hypotheses testing, are related with the Type II errors \cite{Bretz_2011}. In the case of MST identification these error rates are connected with FWER. In particular one has $CPOWER=FWER$. Therefore it does not give a new measure of uncertainty.  

Considered measures of uncertainty  don't take into account the numbers of errors. In multiple hypotheses testing there are error rates which take into account the numbers of errors:  Per-Family Error Rate (PFER), Per-Comparison Error Rate (PCER), Average Power (AVE), or True Positive Rate (TPR). PFER is defined as the expected  number of Type I errors. Associated loss function can be defined as $w_{PFER}=FP$. PCER is defined by $PCER= PFER/M$, $M=N(N-1)/2$.  Loss function for the Average Power (AVE) is defined by $w_{AVE}=(TP/(FN+TP))$. In binary classification $Risk_{AVE}$ is related with True Positive Rate (TPR), or Sensitivity, or Recall. 

For  MST identification all these uncertainty characteristics are related with False Discovery Rate (FDR).  
FDR is defined by the loss function $w_{FDR}=(FP/(FP+TP))$. One has in our case $PFER=(N-1)FDR$, $PCER=2FDR/N$, $AVE=1-FDR$, $TPR=Recall=Precision=1-FDR$. 
In addition, one has 
$$
0 \leq TPR \leq 1, \ \ 0 \leq FPR \leq \frac{FP}{M-(N-1)}=\frac{2}{N-2}.
$$

Another measure of error in binary classification is Accuracy (ACC), or proportion of correct decisions. It is defined  by the following loss function $w_{ACC}=(TP+TN)/M$, $M=N(N-1)/2$. This measure is related with FDR by the formula
$$
ACC= \displaystyle 1- \frac{4FDR}{N}
$$
ACC is not well appropriate for MST  identification because for a large $N$ ACC is close to 1, independently of the number of errors. 

Taking into account the above discussion we argue that FDR is an appropriate measure of uncertainty for MST identification. Note that for MST identification FDR is the proportion of false edges (non correctly identified edges) in MST. 

\section{Kruskal algorithm for MST identification}\label{Kruskal algorithm for MST identification}

Let $(X,\gamma)$ be the random variables network where $X=(X_1,\ldots,X_N)$ be the random vector and $\gamma$ be the pairwise measure of dependence. Let $x_i(t),i=1,\ldots,N;t=1,\ldots,n$ be the observations of $X$ and $\hat{\gamma}_{i,j}$ be the estimations of the $\gamma_{i,j}$ constructed by observations $x_i(t),i=1,\ldots,N;t=1,\ldots,n$, $\hat{\Gamma}=(\hat{\gamma}_{i,j})$. Kruskal  algorithm for MST identification can be described as follows.

\vskip 2mm
\noindent 
{\bf Kruskal algorithm for MST identification by observations:} the Kruskal algorithm calculates the collection of edges $\hat{MST}$ of maximum spanning tree in the network model $(V,\hat{\Gamma})$ by the following steps
\begin{itemize}
\item Sort the edges of the complete weighted graph $(V,\hat{\Gamma})$ into decreasing order by weights $\hat{\gamma}_{i,j}$. 
\item Add the first edge to $\hat{MST}$.
\item Add the next edge to $\hat{MST}$ if and only if it does not form a cycle in the current $\hat{MST}$. 
\item If $\hat{MST}$ has $(N-1)$ edges, stop and output $\hat{MST}$ . Otherwise go to the previous step.
\end{itemize}

Kruskal algorithm for MST identification in Pearson correlation network uses the classical estimations of Pearson correlations (sample Pearson correlations):
\begin{equation}\label{sample_pearson_measure}
 \hat{\gamma}^P_{i,j}=r_{i,j}=\frac{\sum_{t=1}^n(x_i(t)-\overline{x_i})(x_j(t)-\overline{x_j})}{\sqrt{\sum_{t=1}^n(x_i(t)-\overline{x_i})^2\sum_{t=1}^n(x_j(t)-\overline{x_j})^2}} 
\end{equation}
 
\vskip 2mm
Kruskal algorithm for MST identification in Fechner correlation network uses the following estimations of Fechner correlations (sample Fechner correlations):
$$
\displaystyle \hat{\gamma}^{Fh}_{i,j}=2 \hat{\gamma}^{Sg}_{i,j}-1
$$
where $\hat{\gamma}^{Sg}_{i,j}$ are estimations of sign similarities. These estimations are given by
\begin{equation}\label{sample_sign_measure}
\hat{\gamma}^{Sg}_{i,j}=\frac{1}{n}\sum_{t=1}^nI_{i,j}(t) 
\end{equation}
with   
$$
I_{i,j}(t)=\left\{\
							\begin{array}{ll}
							  0,& (x_i(t)-\overline{x_i})(x_j(t)-\overline{x_j})\leq 0\\
								1,& (x_i(t)-\overline{x_i})(x_j(t)-\overline{x_j})>0\\
							\end{array}
						\right.			
$$
where
$$
\overline{x_i}=\displaystyle \frac{1}{n}\sum_{t=1}^n x_i(t), \ \ i=1,2,\ldots,N
$$
In the case when the vector of means $\mu$ is known one can calculate $I_{i,j}(t)$ by
$$
I_{i,j}(t)=\left\{\
							\begin{array}{ll}
							  0,& (x_i(t)-\mu_i)(x_j(t)-\mu_j)\leq 0\\
								1,& (x_i(t)-\mu_i)(x_j(t)-\mu_j)>0\\
							\end{array}
						\right.			
$$
  
\vskip 2mm
Kruskal algorithm for MST identification in Kendall correlation network uses the following estimations of Kendall correlations
\begin{equation}\label{sample_kendall_measure}
 \hat{\gamma}^{Kd}_{i,j}=\frac{1}{n(n-1)}\sum_{t=1}^n\sum_{\begin{array}{l}s=1\\s\neq t \end{array}}^n I^{Kd}_{i,j}(t,s)
\end{equation}
where 
$$
I^{Kd}_{i,j}(t,s)=\left\{\
									\begin{array}{ll}
									  1, & (x_i(t)-x_i(s))(x_j(t)-x_j(s))\geq 0\\
										-1, & (x_i(t)-x_i(s))(x_j(t)-x_j(s))<0
									\end{array}
							 \right.
$$

\section{Robustness of Kruskal algorithm in Fechner correlation network}\label{Robustness of Kruskal algorithm}
Uncertainty of Kruskal algorithm for MST identification depends on the chosen correlation network. From one side, for any $X \in K(\Lambda)$ Kruskal algorithms in different correlation networks  identify the same true MST. From the other side, error in the identification can be different. In this Section we state and prove an interesting property of Kruskal algorithm for MST identification in Fechner correlation network. This property can be associated with robustness of the algorithm. Indeed, robustness in general is associated with non sensitivity of  an algorithm to the change of some parameters. This is the case of Kruskal algorithm for MST identification in Fechner correlation network. More precisely the following statement is true: 
       
\begin{thm}
Let $X \in K(\Lambda)$ and the vector of means $\mu$ be known. Then FDR of Kruscal algorithm for MST identification in Fechner correlation network is the same for any vector $X$. 
\end{thm}
This means that FDR as a risk function does not depend on distribution from the class $K(\Lambda)$ (distribution free risk function).

\noindent
{\bf Proof.} Proof is based on the results from our publication \cite{Kalyagin_2017}. First step of the Kruskal algorithm for MST identification in Fechner correlation network is to sort the edges of the complete weighted graph $(V,\hat{\Gamma}^{Fh})$ into decreasing order by weights $\hat{\gamma}^{Fh}_{i,j}$. 
One has
$$
\displaystyle \hat{\gamma}^{Fh}_{i,j}=2 \hat{\gamma}^{Sg}_{i,j}-1
$$
Therefore, the first step of the algorithm is equivalent to sort $\hat{\gamma}^{Sg}_{i,j}$ in decreasing order. 
It is proved in \cite{Kalyagin_2017} (Theorem 2) that the joint distribution of statistics $\hat{\gamma}^{Sg}_{i,j}$ (in the paper they are denoted by $T^{Sg}_{i,j}$) is the same for any $X \in K(\Lambda)$. It implies that the probability of any ordering of $\hat{\gamma}^{Sg}_{i,j}$ does not depend on distribution of the vector $X \in K(\Lambda)$. $\hat{MST}$, obtained by Kruskal algorithm of identification is completely  defined by ordering of $\hat{\gamma}^{Sg}_{i,j}$. Therefore, any such ordering generates the same numbers FP, FN, TP and TN. It implies that  the distribution of the loss function $w_{FDR}=(FP/(FP+TP))$ is the same for any $X \in K(\Lambda)$ and the theorem follows.

\section{Reliability of Kruskal algorithm in  different correlation networks}\label{Reliability of Kruskal algorithm}

It the section we study by numerical simulations reliability (uncertainty) of Kruskal algorithm for MST identification in three correlation networks: Pearson correlation network, Fechner correlation network, and Kendall correlation network for stock market returns. The results of numerical experiments show that reliability of MST identification is different in three networks, despite the fact that true MST is the same. It is shown  that for Pearson correlation network the FDR of MST identification essentially depends on distribution of stock returns. For Fechner correlation network we observe that the FDR is non sensitive to the assumption on stock's return distribution in accordance with theoretical result of the robustness of Kruskal algorithm. New and surprising phenomena are discovered for   Kendall correlation network. Our experiments show that FDR of Kruskal algorithm for MST identification in Kendall correlation network weakly depend on distributions of the vector $X \in K(\Lambda)$ and at the same time the value of FDR is almost the best in comparison with MST identification in other networks. This needs a further investigation. 

Our experiments are organized as follows. We take the real data of stock returns from a stock market. Using these data we estimate vector of means and  correlation matrix for the stock returns. These estimations are fixed as true vector of means $\mu$ and matrix $\Lambda$ for the random vectors $X$ from the class $K(\Lambda)$ of elliptical distributions (in our experiments, we use correlation matrix as the matrix $\Lambda$). To make our conclusions more general we consider networks of different sizes.  

To study how FDR of Kruskal algorithm for MST identification depends on distribution from the class $K(\Lambda)$ we consider the  family of distributions from this class with the following densities 
$$
f_{\epsilon}(x)=(1-\epsilon) f_{Gauss}(\mu,\Lambda)+\epsilon f_{Student}(\mu,\Lambda), \ \ \epsilon \in[0,1]
$$ 
Here for $f_{Student}(\mu,\Lambda)$ we fix the parameter $\nu=3$. For $\epsilon=0$ we have the multivariate Gaussian distribution, and for $\epsilon=1$ we have the multivariate Student distribution. Other distributions are a mixture of these two distributions. The computational scheme is the following
\begin{itemize}
\item For a given covariance (correlation) matrix $\Lambda$ calculate true MST. 
\item Generate a sample of the size $n$ from distribution with density $f_{\epsilon}(x)$.
\item For each correlation network use Kruskal algorithm to identify MST by observations.
\item Compare true MST and MST identified by Kruskal algorithm and calculate FDR for the sample
\item Repeat last three steps $S$ times and make average of FDR's to evaluate the expected value of FDR. 
\end{itemize}

\begin{figure}[h!]
\centering
\includegraphics*[width=1.00\textwidth]{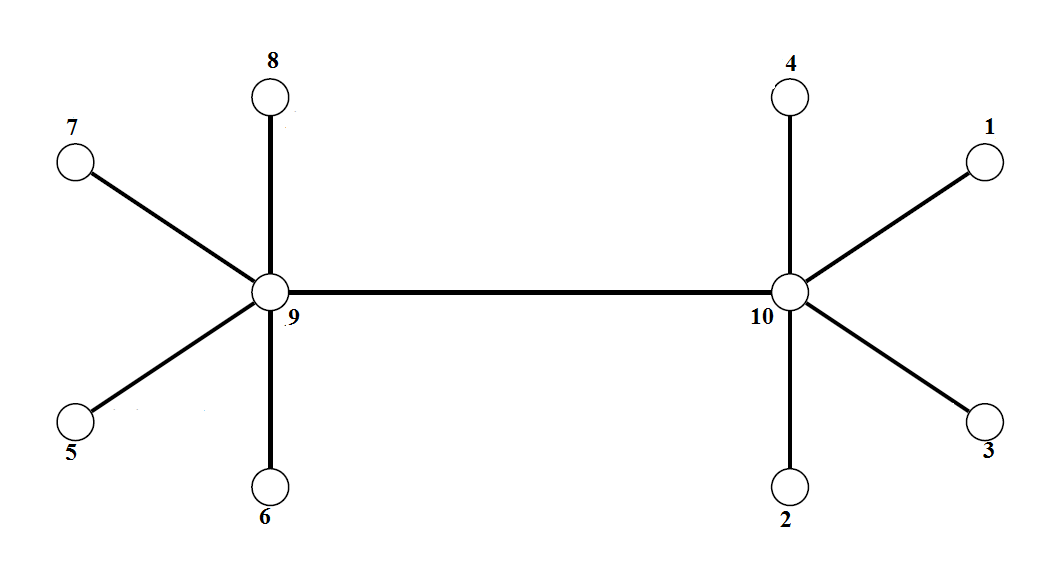}
\caption{True MST, $N=10$}\label{True_MST_N_10}
\end{figure}

\vskip 2mm
\noindent
{Experiment 1.} Consider the following $N=10$ stocks from USA stock market: A (Agilent Technologies Inc), AA (Alcoa Inc), AAP (Advance Auto Parts Inc), AAPL (Apple Inc), AAWW (Atlas Air Worldwide Holdings Inc), ABAX (Abaxis Inc), ABD (ACCO Brands Corp), ABG (Asbury Automotive Group Inc), ACWI (iShares MSCI ACWI Index Fund), ADX (Adams Express Company). We estimate the parameters $\mu$ and $\Lambda$ by the data for the 250 observations started from November 2010. Associated matrix of Pearson correlations is 
$$
\left(
  \begin{array} {cccccccccc} \label{arr:weightMatrix}
               1.0000&    0.7220&    0.4681&    0.4809&    0.6209&    0.5380&    0.6252&    0.6285&    0.7786&    0.7909\\
           0.7220&    1.0000&    0.4395&    0.5979&    0.6381&    0.5725&    0.6666&    0.6266&    0.8583&    0.8640\\
            0.4681&    0.4395&    1.0000&    0.3432&    0.3468&    0.2740&    0.4090&    0.4016&    0.4615&    0.4832\\
            0.4809&    0.5979&    0.3432&    1.0000&    0.4518&    0.4460&    0.4635&    0.4940&    0.6447&    0.6601\\
            0.6209&    0.6381&    0.3468&    0.4518&    1.0000&    0.5640&    0.5994&    0.5369&    0.7170&    0.7136\\
           0.5380&    0.5725&    0.2740&    0.4460&    0.5640&    1.0000&    0.4969&    0.4775&    0.6439&    0.6242\\
            0.6252&    0.6666&    0.4090&    0.4635&    0.5994&    0.4969&    1.0000&    0.6098&    0.7161&    0.7158\\
           0.6285&    0.6266&    0.4016&    0.4940&    0.5369&    0.4775&    0.6098&    1.0000&    0.6805&    0.6748\\
            0.7786&    0.8583&    0.4615&    0.6447&    0.7170&    0.6439&    0.7161&    0.6805&    1.0000&    0.9523\\
            0.7909&    0.8640&    0.4832&    0.6601&    0.7136&    0.6242&    0.7158&    0.6748&    0.9523&    1.0000
        \end{array}
    \right)
$$
True MST, obtained from this matrix is given by Fig. \ref{True_MST_N_10}. 

The results of FDR evaluation for Kruskal algorithm of MST identification in three networks are presented in Tables \ref{FDR_N10_n10},\ref{FDR_N10_n100},\ref{FDR_N10_n1000}. Analysis of the results shows that for $n=10$ and $n=100$ all algorithms for MST identification have weak reliability in terms of FDR. The results of Table \ref{FDR_N10_n1000} show that 1-2 edges in identified MST are different from edges in true MST. 

Interesting results were obtained for Kendall correlation network. Namely the obtained results shows that quality of MST identification in Kendall correlation network is close to the quality of MST identification in Pearson network for Gaussian distribution and are better than obtained results of MST identification in Pearson network for Student distribution. This is valid for $n=1000$ too. Besides one can see strong dependence on distribution of FDR for Kruskal algorithm of MST identification in Pearson correlation network and stability of FDR for Kruskal algorithm of MST identification in Fechner correlation network.  From the other side, FDR for Kruskal algorithm of MST identification in Kendall correlation network is almost stable, weakly depending on distribution. 
 
\vskip 2mm
\noindent
{Experiment 2.} Consider $N=50$ stocks from NASDAQ stock market with largest trade volume for the year 2014. Parameters $\mu$ and $\Lambda$ are estimated by  250 observations for 2014. 
The tables \ref{FDR_N50_n1000} and \ref{FDR_N50_n10000} present the results of FDR evaluation for $n=1000$ and $n=10000$ for three networks. The results are almost the same as for Experiment 1. One can see strong dependence on distribution of FDR for Kruskal algorithm of MST identification in Pearson correlation network and stability of FDR for Kruskal algorithm of MST identification in Fechner correlation network.  From the other side, FDR for Kruskal algorithm of MST identification in Kendall correlation network is almost stable, weakly depending on distribution. Reliability of MST identification in this network is almost the best with respect to other networks.

%As true MST the MST  

\begin{table}[h!]
\begin{center}
  \begin{tabular}{|c|c|c|c|c|c|c|c|c|c|c|c|}
\hline
 measure,$\epsilon $ & 0    & 0.1  & 0.2  & 0.3  & 0.4  & 0.5  & 0.6  & 0.7  & 0.8  & 0.9  & 1    \\ \hline
 Pearson 						 & 0.66 & 0.67 & 0.67 & 0.67 & 0.68 & 0.68 & 0.67 & 0.68 & 0.69 & 0.68 & 0.69 \\ \hline
 Fechner             & 0.65 & 0.64 & 0.64 & 0.64 & 0.64 & 0.64 & 0.64 & 0.64 & 0.64 & 0.63 & 0.64 \\ \hline
 Kendall             & 0.65 & 0.65 & 0.66 & 0.66 & 0.66 & 0.67 & 0.67 & 0.67 & 0.67 & 0.66 & 0.66 \\ \hline

  \end{tabular}
\end{center}
 \caption{False discovery rate. N=10, n=10}\label{FDR_N10_n10}
 \end{table}

\begin{table}[h!]
\begin{center}
  \begin{tabular}{|c|c|c|c|c|c|c|c|c|c|c|c|}
\hline
 measure,$\epsilon $ & 0    & 0.1  & 0.2  & 0.3  & 0.4  & 0.5  & 0.6  & 0.7  & 0.8  & 0.9  & 1    \\ \hline
 Pearson 						 & 0.37 & 0.40 & 0.40 & 0.41 & 0.43 & 0.45 & 0.46 & 0.48 & 0.48 & 0.50 & 0.52 \\ \hline
 Fechner             & 0.52 & 0.53 & 0.54 & 0.53 & 0.53 & 0.53 & 0.53 & 0.54 & 0.53 & 0.53 & 0.53 \\ \hline
 Kendall             & 0.41 & 0.40 & 0.41 & 0.41 & 0.42 & 0.42 & 0.42 & 0.44 & 0.44 & 0.44 & 0.44 \\ \hline

  \end{tabular}
\end{center}
 \caption{False discovery rate. N=10, n=100}\label{FDR_N10_n100}
 \end{table}

\begin{table}[h!]
\begin{center}
  \begin{tabular}{|c|c|c|c|c|c|c|c|c|c|c|c|}
\hline
 measure,$\epsilon $ & 0    & 0.1  & 0.2  & 0.3  & 0.4  & 0.5  & 0.6  & 0.7  & 0.8  & 0.9  & 1    \\ \hline
 Pearson 						 & 0.15 & 0.17 & 0.19 & 0.21 & 0.22 & 0.23 & 0.26 & 0.29 & 0.30 & 0.33 & 0.34 \\ \hline
 Fechner             & 0.34 & 0.34 & 0.33 & 0.33 & 0.33 & 0.33 & 0.33 & 0.33 & 0.33 & 0.34 & 0.33 \\ \hline
 Kendall             & 0.17 & 0.17 & 0.17 & 0.17 & 0.18 & 0.18 & 0.18 & 0.18 & 0.19 & 0.20 & 0.20 \\ \hline

  \end{tabular}
\end{center}
 \caption{False discovery rate. N=10, n=1000}\label{FDR_N10_n1000}
 \end{table}

\begin{table}[h!]
\begin{center}
  \begin{tabular}{|c|c|c|c|c|c|c|c|c|c|c|c|}
\hline
 measure,$\epsilon $ & 0    & 0.1  & 0.2  & 0.3  & 0.4  & 0.5  & 0.6  & 0.7  & 0.8  & 0.9  & 1    \\ \hline
 Pearson 						 & 0.23 & 0.26 & 0.29 & 0.34 & 0.37 & 0.34 & 0.42 & 0.44 & 0.46 & 0.50 & 0.52 \\ \hline
 Fechner             & 0.41 & 0.39 & 0.40 & 0.40 & 0.39 & 0.40 & 0.40 & 0.41 & 0.41 & 0.40 & 0.41 \\ \hline
 Kendall             & 0.25 & 0.25 & 0.25 & 0.26 & 0.27 & 0.27 & 0.28 & 0.27 & 0.28 & 0.29 & 0.28 \\ \hline

  \end{tabular}
\end{center}
 \caption{False discovery rate. N=50, n=1000}\label{FDR_N50_n1000}
 \end{table}

\begin{table}[h!]
\begin{center}
  \begin{tabular}{|c|c|c|c|c|c|c|c|c|c|c|c|}
\hline
 measure,$\epsilon $ & 0    & 0.1  & 0.2  & 0.3  & 0.4  & 0.5  & 0.6  & 0.7  & 0.8  & 0.9  & 1    \\ \hline
 Pearson 						 & 0.08 & 0.10 & 0.13 & 0.14 & 0.16 & 0.18 & 0.21 & 0.25 & 0.24 & 0.28 & 0.32 \\ \hline
 Fechner             & 0.14 & 0.13 & 0.14 & 0.13 & 0.14 & 0.14 & 0.14 & 0.14 & 0.13 & 0.13 & 0.14 \\ \hline
 Kendall             & 0.09 & 0.08 & 0.08 & 0.09 & 0.08 & 0.08 & 0.09 & 0.09 & 0.09 & 0.08 & 0.10 \\ \hline

  \end{tabular}
\end{center}
 \caption{False discovery rate. N=50, n=10000}\label{FDR_N50_n10000}
 \end{table}

\section{Concluding remarks}\label{Concluding remarks}

The main advantage of the proposed framework to measure uncertainty of algorithms for MST identification is that it allows to make a correct comparison of the uncertainty for different networks and for a large class of distributions. Peculiarities of Pearson, Fechner and Kendall correlation networks for elliptical distributions were emphasized in the paper on the base of this approach. It was observed that Kendall correlation network looks the most appropriate for MST identification. This phenomena will be a subject for further investigations.


\begin{thebibliography}{99}

\bibitem{Anderson_2003} Anderson T.W. An introduction to multivariate statistical analysis.3-d edition. Wiley-Interscience, New York., 2003.

\bibitem{Boginski_2006}     Boginski V., Butenko S., Pardalos P.M.    Mining market data: a network approach, J. Computers and Operations Research. 33 (11) 3171--3184 (2006).

\bibitem{Bretz_2011} Bretz F., Hothorn T., Westfall P. Multiple Comparisons Using R. Taylor and Francis Group, 2011.

\bibitem{Gross_2006} Gross J. and Yellen J. Graph Theory and Its Applications. CRC Press, 2006.

\bibitem{Gupta_2013} Gupta F. K., Varga T., and Bodnar T. Elliptically Contoured Models in Statistics and Portfolio Theory. Springer, 2013.

\bibitem{Fang_1990} Fang K.T. Kotz S. Ng K.W. Symmetric multivariate and related distributions, Chapman and Hall, London, 1990.  

\bibitem{Fang_2002} Fang H.B. Fang K.T. The Meta-elliptical Distributions with Given Marginals, Journal of Multivariate Analysis, v. 82 (2002), p.1-16

\bibitem{Hochberg_1987} Hochberg Y. and Tamhane A.C. Multiple Comparison Procedures, John Wiley and Sons, New
York, 1987.

\bibitem{Kalyagin_2017} Kalyagin V., Koldanov A., Koldanov P. Robust identification in random variable networks, Journal of Statistical Planning and Inference, Volume 181 (2017), Pages 30-40

\bibitem{Kalyagin_2020} Kalyagin V. A., Koldanov A. P.,  Koldanov P.A., Pardalos P.M. Statistical analysis of graph structures in random variable networks, Springer Brief in Optimization,  Springer, 2020.  

\bibitem{Kalyagin_2014_1} Kalyagin V.A., Koldanov A.P., Koldanov P.A., Pardalos P.M., Zamaraev V.A.        Measures of uncertainty in market network analysis, Physica A: Statistical Mechanics and its Applications, v. 413,  No. 1, pp. 59-70 (2014).

\bibitem{Kruskal_1958} Kruskal W.H. Ordinal measures of association. Journal of American Statistical Association, 53:814–861, 1958.

\bibitem{Lindskog_2003} Lindskog F., McNeil A., Schmock U.  Kendalls Tau for Elliptical Distributions. In: Bol G., Nakhaeizadeh G., Rachev S.T., Ridder T., Vollmer KH. (eds) Credit Risk. Contributions to Economics. Physica-Verlag HD, pages 149–156, 2003.

\bibitem{Lehmann_2005} Lehmann E. L. and Romano J. P. Testing Statistical Hypotheses. Springer, 3rd edition, 2005.

\bibitem{Marti_Arxive} Marti G. Nielsen F.  Bińkowski M. Donnat P. A review of two decades of correlations, hierarchies, networks and clustering in financial markets,  arXiv:1703.00485v4

\bibitem{Sensoya_2014} Sensoya A., Tabak B.M. Dynamic spanning trees in stock market networks: the case of AsiaPacific. Phys. A Stat. Mech. Appl. 414:387–402 (2014)

\bibitem{Tumminello_2007}  Tumminello M.  Coronello C. Lillo F. Micciche S.  Mantegna G. Spanning tree and bootstrap reliability estimation in correlation-based network, International Journal of Bifurcation and ChaosVol. 17 (2007), No. 07, pp. 2319-2329 

\bibitem{Walle_2018} Valle, M.A., Ruz, G.A., Morris, R. Market basket analysis: Complementing association rules with minimum spanning trees, Expert Systems with Applications, v. 97 (2018), p. 146-162  

\bibitem{Wang_2012} Wang, G.-J., Chi, X., Han, F., Sun, B.: Similarity measure and topology evolution of foreign
exchange markets using dynamic time warping method: evidence from minimal spanning tree.
Phys. A Stat. Mech. Appl. 391(16):4136–4146 (2012)

\end{thebibliography}
\end{document}